\documentclass[12pt]{iopart} \usepackage{epsfig}
        \usepackage{amssymb}
\begin{document}
         \title[On demand entanglement in DQDs]{On demand entanglement
           in double quantum dots via coherent carrier scattering}
         \author{F Buscemi$^{1,2}$, P Bordone$^{3,4}$ and A
           Bertoni$^4$}

         \address{$^1$ Department of Electronics Computer Science and
           Systems University of Bologna, Viale Risorgimento 2,
           I-40136 Bologna, Italy}

         \address{$^2$ ARCES, Alma Mater Studiorum, University of
           Bologna, Via Toffano 2/2, 40125 Bologna, Italy}

          \address{$^3$ Department of Physics, University of Modena and
           Reggio Emilia, Via Campi 213/A,I-41125 Modena, Italy}
	\address{$^4$ Centro S3, CNR-Istituto  Nanoscienze, Via
           Campi 213/A, I-41125, Modena, Italy }
         \eads{\mailto{fabrizio.buscemi@unimore.it} }

         \date{\today}
         \begin{abstract}
           We show how two qubits encoded in the orbital states of two
           quantum dots can be entangled or disentangled in a
           controlled way through their interaction with a weak electron
           current.  The transmission/reflection
           spectrum of each scattered electron, acting as an
           entanglement mediator between the dots, shows a signature
           of the dot-dot entangled state.  Strikingly, while few
           scattered carriers produce decoherence of the whole
           two-dots system, a larger number of electrons injected from
           one lead with proper energy is able to recover its quantum
           coherence.  Our numerical simulations are based on a
           real-space solution of the three-particle Schr\"odinger
           equation with open boundaries. The computed transmission
           amplitudes are inserted in the analytical expression of the
           system density matrix in order to evaluate the
           entanglement.

         \end{abstract}
         \submitto{\NJP}
         \section{Introduction}
         Among the various proposals and schemes advanced for
         realiable quantum computing
         architectures~\cite{Kane,Naka,Loss,Fedi}, semiconductor
         double quantum dots (DQDs) are considered very
         promising candidates for the realization of quantum bits and
         gates~\cite{Loss,Fedi,Div,Brandes,Wu}. Indeed, these structures can arbitrarily be scaled to
         large systems and could be easily integrable with other
         microlectronic devices. Besides their potentialities in the
         frame of quantum information science, DQDs are also very
         interesting from the basic physics point of view, as they
         enable both to analyze the peculiar features of electron
         transport phenomena, and to relate them to the appearance of
         quantum correlations~\cite{Wiel, vanWees, Meirav, Lassen}.

         A number of implementations of DQDs qubits have been
         investigated from the theoretical and experimental points of
         view~\cite{Loss,Fedi,Div,Brandes,Wu}. Two degrees of freedom,
         spin and charge, can be used to encode the qubit.  While the
         feasibility of quantum logic gates acting on spin states is
         hampered by the need of local magnetic fields~\cite{Div2},
         state-of-the-art nanofabrication technology~\cite{Haya,Petta,Shink}
         allows for a precise control of the local charge and orbital  degrees  of
         freedom.  Recently, Shinkay \emph{et al.}~\cite{Shink} have
         realized the coherent manipulation of charge states
         in two spatially separated DQDs integrated in a GaAs/AlGaAs
         heterostructure.  Specifically, multiple two-qubit
         operations, such as the controlled-rotation and the swap,
         have been successfully implemented.

         The main threat to the correct functioning of quantum
         information processing devices is represented by the
         decoherence stemming from the interaction with the external
         environment or, from a different perspective, by the
         uncontrolled entanglement of the qubits with the environment.
         For charge states in semiconductor DQDs, the loss of
         coherence is mainly due to the coupling of the carriers to
         the crystal lattice vibrations and to the Coulomb interaction
         with other charged particles~\cite{Stavrou}. In fact, the
         decoherence induced by the electron-phonon interaction has
         been widely investigated in the literature~\cite{Wu,
           Voro,Cao,Openov,Li}, where the quantum dots (QDs) are
         usually considered as two point-like systems with two energy
         levels coupled to the phonon bath. Analytical estimations of
         the decoherence effects and of their characteristic timescales 
         have been given by using various techniques, ranging
         from the Born-Markov approximation~\cite{Voro} to
         perturbative calculations in a non-Markovian
         regime~\cite{Wu,Cao}.  On the other hand, the role played by
         the Coulomb interaction between charge carriers into the loss
         of coherence has not been deeply analyzed.

         In this paper, we intend to investigate the entanglement
         properties of bound electrons in a GaAs DQD when other
         electron pass through the structure.  In particular, we focus
         on the appearance of quantum correlations in a three-particle
         scattering, where charged particles incoming from a lead
         enter, one at a time, in a DQD structure and interact via
         Coulomb potential  with two electrons, one in each dot.  In
         our scheme, the electrons crossing the device have the double
         role of entanglement ``mediator'' between the two dots and
         between the DQDs system and the leads, i.e. the environment.
         The aim of this work is to show how the Coulomb interaction
         between the system and the ``mediator'' can be, under certain
         conditions, a suitable means to entangle or disentangle in a
         controlled way the qubit states encoded in the
         single-particle energy levels of the dots. A detailed
         theoretical estimation of such effects results to be of
         interest  also for the experimental feasibility of the above
         qubit, as the results reached in the production,
         manipulation, and coherent control of charge states in DQDs
         seem to indicate~\cite{Haya,Petta,Shink}.  In this view the
         entanglement/disentanglement of the DQDs can be connected to
         the precise engineering or the suitable tuning of physical
         and geometrical parameters, such as the DQDs level spacing or
         the current intensity, here modelled as the successive
         injection of carriers in the scattering region.

         The numerical procedure used to solve the model is a
         generalization of the quantum transmitting boundary method
         (QTBM)~\cite{lent,bertoni}.  It allows us to find the
         reflection and transmission amplitudes of each scattering
         channel as a function of the initial kinetic energy of the
         incoming carrier and of the state occupation of the dots. Our
         analysis is time-independent, in the sense that the
         few-particle scattering states are obtained by the solution
         of a time-independent open-boundary Schr\"odinger equation,
         and therefore does not permit to evaluate the dynamics and
        the  characteristic timescales of quantum
         correlations~\cite{bertoni,busce}. On the other hand, the
         QTBM takes explicitely into account the spatial structure and
         therefore the size and the shape of the dots, thus permitting
         to overcome the approximations implied in the description of
         a dot in terms of a two-level point-like system. We
         single-out the peculiar mechanisms of electron transport
         through DQDs resulting in resonances in the transmission and
         reflection spectra and thus leading to entanglement and
         decoherence~\cite{Oliver,Lopez,busce,Kazu,Cic1}.  In the
         evaluation of such effects, both the transmitted and reflected
         components of the scattered wavefunction are taken into
         account.
       
         The paper is organized as follows. In
         Section~\ref{PhysicalSys}, we introduce the physical system
         reproducing a DQD structure in GaAs, and illustrate the
         computational approach adopted to find the few-particle
         scattering states.  The description of the DQD in terms of a
         two-qubit model and the discussion about the theoretical
         procedures used to evaluate the entanglement and the
         decoherence are given in Section~\ref{Twoqu}.  In
         Section~\ref{Results}, we show the numerical results first
         obtained for the scattering of a single carrier, then for a
         weak electric current. For the latter case, the condition leading
        to  the
         maximum entanglement production or to the complete
         disentanglement are analyzed in detail. Finally, in
         Section~\ref{Conclu} we comment on the results and draw our
         conclusions.

         \section{The DQD structure and the computational
           approach}\label{PhysicalSys}

\begin{figure}[h]
           \begin{center}
             \includegraphics*[width=0.7\linewidth]{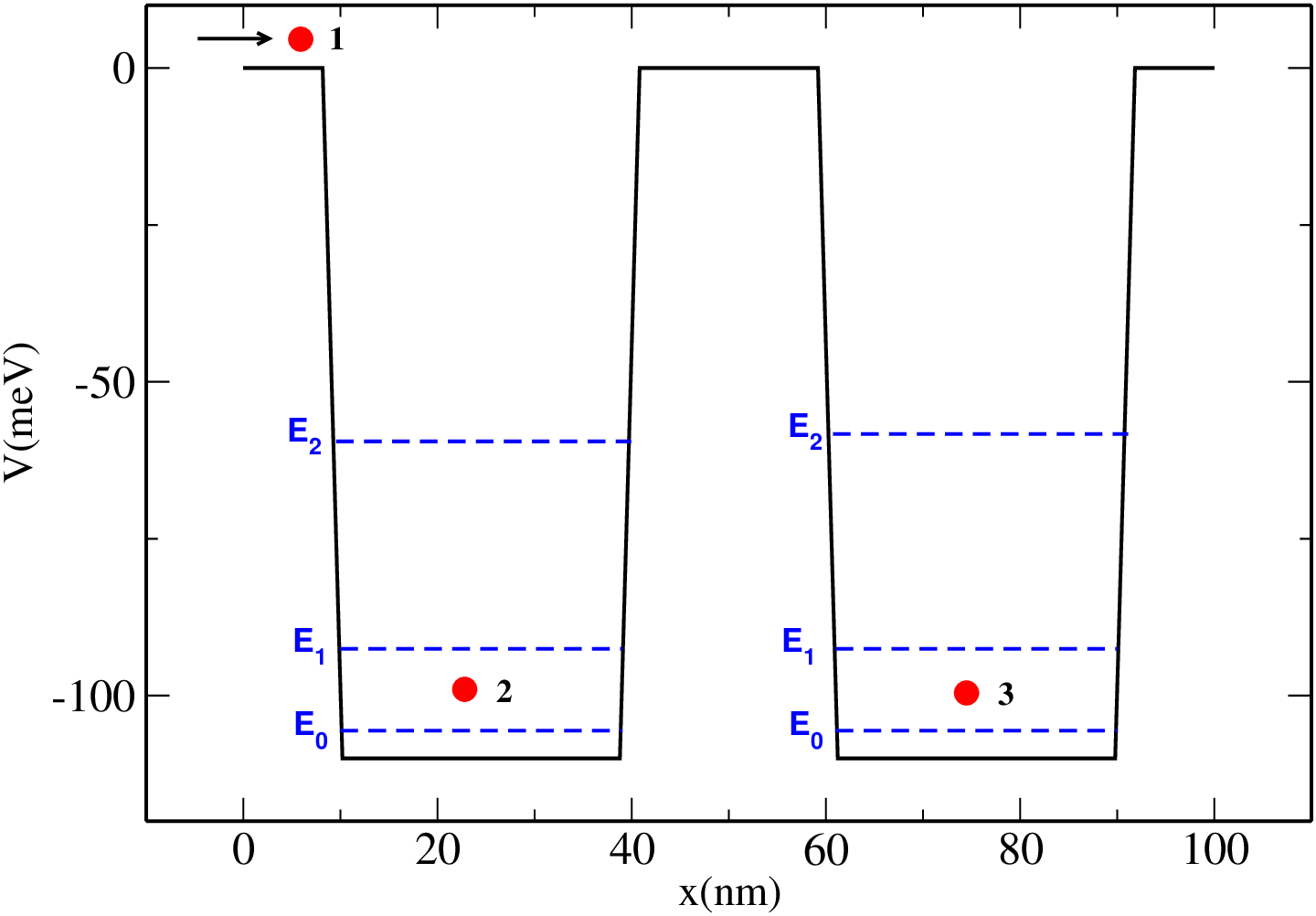}
             \caption{\label{fig1} Profile of the potential $V_s(x)$
               in the scattering region of length $L$=100 nm: the two
               potential wells are 110 meV deep and 30 nm wide and are
               separated by a 20 nm barrier. The dashed lines indicate
               the single-particle energy levels $E_0$, $E_1$, and $E_2$ of
               the ground, the first and second excited state of the dots
               respectively, with $E_0$=-105.6 meV,$E_1$=-92.5,  and $E_2$=-71.5
               meV. With our parameters the levels $E_2$ (and above) are found to have negligible occupancy.
               In our numerical calculations we take $m^{\ast}$=0.067 $m_e$,
               with  $ m_e$ indicating the bare mass electron, and $\varepsilon$=12.9.}
           \end{center}
         \end{figure}

         In our model (see Figure~\ref{fig1}), we consider an electron incoming from the left,
         with respect to the DQD one dimensional (1D) structure, with kinetic energy $T_0$.  We
         examine the case of one scattered carrier at a time, i.e.  we
         suppose that an electron enters the scattering region only
         after the previous one has already left.  Such an assumption,
         and the fact that the charging energy of the DQD is larger
         than the spacing between the two-particle energy levels, means
         that our system always operates in three-particle regime.
         The incident particle is scattered via Coulomb interaction by
         the two electrons bound in a structure potential $V_s$ in a
         region of length $L$ (Figure~\ref{fig1}) mimicking a DQD
         structure and constituted by two potential wells separated by
         a potential barrier wide enough to make negligible the
         Coulomb interaction between the two confined particles.
         Moreover, the structure is connected to external leads kept
         at zero potential.  The $N$ two-particle bound states and
         energies of the DQD will be indicated by $|\Xi_n \rangle $
         and $\epsilon_n$, respectively (with $n$0 indicating
         the ground state). As it will be shown in
         the following, some of them can also be expressed in terms of
         $|\chi_l^{R} \chi_m^{L}\rangle $ and $(E_l^{R}+E_m^{L})$,
         where $|\chi_l^{R} \rangle (|\chi_m^{L}\rangle)$ indicates
         the single-particle bound state of the right (left) dot with
         energy $E_l^{R}(E_m^L)$. Due to the symmetry of the potential
         $V_s$, $E_l^{R}=E_l^L$.

         As anticipated, we restrict our investigation to a 1D
         analysis.  Such an assumption, needed to solve numerically 
        the few-particle problem
         with open boundaries, is physically reasonable if the
         transverse dimension of the structure is small compared with
         the other length scales. In this case, all the particles can
         be supposed to occupy the lowest single-particle transverse
         subband~\cite{Fogler}.  Furthermore, we consider for the
         incoming electrons only energies below the ionization
         threshold of the DQD. This means that when the outgoing electrons
         leave the scattering region, either reflected or transmitted,
         the confined particles remain in a bound state of the DQD.

         The three-particle Hamiltonian $\mathcal{H}$ is given by
         \begin{equation} \label{Ham}
           \mathcal{H}(x_1,x_2,x_3)=\mathcal{H}_0(x_1)+\mathcal{H}_0(x_2)+\mathcal{H}_0(x_3)+\sum_{i=1}^3
           \sum_{j=1}^{i-1}\frac{e^2 }{4\pi \varepsilon r_{ij}}\exp{(-
             r_{ij}/\lambda_d)},
         \end{equation}
         where $r_{ij}=\sqrt{\left(x_i-x_j\right)^2+d^2}$ and
         $\mathcal{H}_0(x_i)$ is the single-particle Hamiltonian
         \begin{equation}
           \mathcal{H}_0(x_i)=-\frac{\hbar^2}{2m^{\ast}} \frac{\partial^2}{\partial x_i^2} + V_s(x_i).
         \end{equation}
         $\varepsilon $ and $m^{\ast}$ are the dielectric constant and
         effective mass of the GaAs, respectively. The term describing
         the mutual interaction between particles in the r.h.s. of
         equation~(\ref{Ham}) is a screened Coulomb potential with a
         Debye length $\lambda_d$, here taken significantly larger
         than the characteristics length of the structure.
         Furthermore, the Coulomb term also accounts for the
         transversal dimension $d$ of the confined system (with $d$=1
         nm) through a cut-off term.  In our approach, the fermionic
         nature of the carriers is explicitly accounted for by
         antisymmetrizing the two-  and three-particle
         wavefunctions  $\Xi_n(x_2,x_3)$ and $\Psi(x_1,x_2,x_3)$ for any two-particle
         exchange. It is worth noting that the Hamiltonian given in
         equation~(\ref{Ham}) does not include spin-orbit terms. As a
         consequence, since the orbital wavefunction is
         antisymmetric, we are simulating a three-particle system with
         a symmetric spin component, as in the case of three spin-up
         (or spin-down) electrons. Furthermore, we do not include
         electron-phonon interaction. Infact, the aim of our work is
         to investigate the role played by the Coulomb interaction
         among the electrons  in the system into the
         appearance of entanglement and decoherence.
         Our device is supposed to operate 
         at a temperature in mK regime, where only spontaneous emission is effective 
         and ignoring the coupling of the electrons with the surrounding crystal lattice does not constitute
         a crucial approximation, as we will describe in the following. 
         In GaAs-based structures with levels  splitting of few meV, single-phonon processes 
         lead to an excited-state lifetime of the order of 10$^{-10}$ s, 
          while multi-phonon and multi-electron processes are  orders of magnitude less frequent 
          (see References~\cite{Hu,Bertons}). 
           Given a kinetic energy of 15 meV,
           a single electron traverse our 100 nm long device in about 10$^{-13}$ s.
           If we suppose independent electrons injected from the lead at a mean rate of
             one every 10$^{-12}$ s, corresponding to a current of about 0.16 $\mu$A, 
             100 carriers can be scattered through the double-dot  before a phonon-induced 
           relaxation takes place.  As a consequence, in the following sections
         where we consider a weak electric current, we will limit our calculation up to 60 electrons.

         The scattering states of the three particles are obtained by
         solving the time-independent open-boundary Schr\"odinger
         equation $\mathcal{H} \Psi=E \Psi$ in the cubic domain
         $\big\{x_1,x_2,x_3\big\}$ with $x_i\in [0,L]$ and with the Hamiltonian $\mathcal{H}$
         given in equation~(\ref{Ham}). For this purpose, we applied a
         few-particle generalization of the so-called
         QTBM~\cite{lent}, allowing one to include proper
         open-boundary conditions for each edge of the domain. These
         describe the particle coming from the left as a plane wave
         with energy $T_0$ and wavevector $k_0$
         while the other two electrons are set in a
         two-particle bound state  $|\Xi_j \rangle $ of the DQD  with
         energy $\epsilon_j$.  Moreover, to account for the exchange
         symmetry of the three-particle wavefunction, also the
         antisymmetry of the boundary conditions is imposed, as shown
         in previous works~\cite{bertoni,busce}.

         The correlated scattering state when the particle $1$ is
         localized in the left lead (that is $x_1<0$) reads
         \begin{eqnarray} \label{refl} \fl \Psi(x_1,x_2,x_3)
           \bigg|_{(x_1<0)}= \nonumber \\\fl \Xi_j(x_2,x_3)
           e^{ik_0x_1}+\sum_{n=0}^M b_{jn} \Xi_n(x_2,x_3)
           e^{-ik_{(j-n)}x_1}+\sum_{n=M+1}^{\infty} b_{jn}
           \Xi_n(x_2,x_3) e^{k_{(j-n)}x_1}
         \end{eqnarray}
         where $j=0\ldots N$ is the index of the initial two-particle  DQD state.
         Analogously, when the particles 2 and 3 are in the left lead,
         the boundary conditions are $\Psi(x_1,x_2,x_3)
         |_{(x_2<0)}$=$-\Psi(x_2,x_1,x_3)| _{(x_2<0)}$ and
         $\Psi(x_1,x_2,x_3) |_{(x_3<0)}$=
         $\Psi(x_3,x_1,x_2)|_{(x_3<0)}$, respectively. In the above
         expression $k_{(j-n)}$=$\sqrt{2m^{\ast}T_{(j-n)}}$, where
         $T_{(j-n)}$ denotes the kinetic energy of an electron freely
         propagating in the lead, as obtained by energy conservation
         $T_{(j-n)}=T_0+\epsilon_j-\epsilon_n$. For sake of simplicity,
         we set $\hbar$=1 here and in the following.

         The first term appearing in the r.h.s of
         equation~(\ref{refl}) describes electron 1 incoming from the
         left lead as a plane wave with energy $T_0$, while the other
         electrons are in the two-particle bound state
         $|\Xi_j\rangle$. The second term represents the linear
         combination of all the energetically-allowed possibilities
         when particle 1 is reflected back as a plane wave with wave
         vector $k_{(j-n)}$ and the DQD is in the state
         $|\Xi_n\rangle$. $M$ indicates the number of states for which
         $T_{(j-n)} \ge 0$. The last term accounts for those states
         with $T_{(j-n)}$ negative, which describe particle 1 as an
         evanescent wave in the left lead.  Therefore, the
         coefficients $b_{jn}$ are the transition amplitudes between
         the initial state $\Xi_j(x_2,x_3) e^{ik_0x_1}$ and the final
         state $\Xi_n(x_2,x_3) e^{ik_{(j-n)}x_1}$ when the incoming
         carrier is reflected.

         If particle 1 is in the right lead, the three-particle
         wavefunction takes a form similar to expression~(\ref{refl}),
         which describes the outgoing travelling and evanescent modes
         of the electron in the right lead ($x_1>L$):
         \begin{eqnarray} \label{trasm} \fl \Psi(x_1,x_2,x_3)
           \bigg|_{(x_1>L)}= \sum_{n=0}^M c_{jn} \Xi_n(x_2,x_3)
           e^{ik_{(j-n)}x_1}+\sum_{n=M+1}^{\infty} c_{jn}
           \Xi_n(x_2,x_3) e^{-k_{(j-n)}x_1},
         \end{eqnarray}
         while $\Psi(x_1,x_2,x_3) |_{(x_2>L)}$=$-\Psi(x_2,x_1,x_3)|
         _{(x_2>L)}$ for $x_2>L$ and $\Psi(x_1,x_2,x_3)
         |_{(x_3>L)}$=\\$\Psi(x_3,x_1,x_2)|_{(x_3>L)}$ for $x_3>L.$
         The coefficients $c_{jn}$ describe the transition amplitudes
         in the $n$-th channel, i.e. when the bound particles are in
         the $n$-th state of the DQD. The boundary
         conditions are given by equations~(\ref{refl}) and (\ref{trasm}) with $x_1$=0 and $x_1$=$L$.
          They are coupled to the
         Schr\"odinger equation and discretized by a
         finite-difference method. A system of seven equations is
         obtained, whose numerical solution provides the unknown
         coefficients $b_{jn}$ and $c_{jn}$, and the three-particle
         wavefunction $\Psi(x_1,x_2,x_3)$ in the internal points.

	\section{Decoherence and entanglement of the two-qubit model}\label{Twoqu}
        As a consequence of the scattering, quantum correlations
        between the single-particle energy levels of the bound
        electrons and the energies of the scattered electron appear.
        They are responsible both for the loss of quantum coherence of
        the two-particle state of  the DQD and for the
        building up of quantum entanglement between the two dots.
        First, we show that under some approximations the DQD system
        can be reduced to a two-qubit model. Then, we describe in detail
        the theoretical approach used to evaluate entanglement and
        decoherence in such a model. 
	 
        Although the fermionic nature of the carriers has been explicitly
        taken into account, as shown in
        Section~\ref{PhysicalSys}, in order to solve numerically the physical
        system, we do not use entanglement criteria for
        identical particles~\cite{Sch, Ghira, busce3}. In fact, the
        scattered carrier, either transmitted or reflected, can be
        assumed to be far from the scattering region, while the bound
        particles are essentially trapped into two deep potential
        wells far away from each other. So the spatial overlap between
        the particles results to be negligible. Therefore, the position
        variables can be used to distinguish the particles, while
        quantum correlations are evaluated between scattering channels
        \cite{Eckert,busce}. By moving from spatial to energy
        representation for the quantum states, and taking as input
        state $|\Phi_{IN}\rangle$=$|T_0 \epsilon_j \rangle$, which
        describes the carrier incoming from the left lead with kinetic
        energy $T_0$ and the particles bound with energy $
        \epsilon_j$, the output $|\Phi_{OUT}\rangle$ reads
        \begin{eqnarray}\label{phiout}
          |\Phi_{OUT}\rangle=\sum_{n=0}^M \tilde{b}_{jn}|T_{(j-n)}^< \epsilon_n \rangle+\sum_{n=0}^M \tilde{c}_{jn} |T_{(j-n)}^> \epsilon_n \rangle
        \end{eqnarray}
        where the coefficients $\tilde{b}_{jn}$ are given by
        $\tilde{b}_{jn}=b_{jn}/(\sum_{n=0}^{M}(|b_{jn}|^2+|c_{jn}|^2)$,
        and the analogous expression holds for $\tilde{c}_{jn}$, while
        $|T_{(j-n)}^{<(>)} \epsilon_n \rangle$ indicates the state
        with the carrier reflected (transmitted) as a  plane wave (with kinetic
        energy $T_{(n-j)}= k_{(n-j)}^2/2m$) and the
        other two electrons bound in $|\Xi_n\rangle$ with energy $
        \epsilon_n$.  It is worth noting that in the above expression
        we have omitted the reflected and transmitted outgoing
        evanescent modes, since their contribution to the total
        current is  zero and cannot be
        responsible of any entanglement.

        So far, we considered   the case of
        the injection of a single carrier in the scattering region
        when the two particles trapped in the DQD structure are
        described by a pure state. Let us now examine the injection of
        a second electron also with kinetic energy $T_0$  in the scattering region,
        occurring  after the exit of the previous one from the DQD structure
        via transmission or reflection. We indicate as $a$ and $b$ the first and the second injected electron, respectively. When $b$ enters the DQD,
       the  bound electrons  are not in
        a two-particle pure state, since  they are coupled to the energy
        levels of the carrier $a$, as shown by
        equation~(\ref{phiout}). Therefore, the scattering between the
        electron $b$ and the other two bound in the DQD, will give
        the four-particle state
        \begin{eqnarray}\label{phiout2}
          \fl   |\Phi_{OUT}^{(b,a)}\rangle=\sum_{n=0}^M \sum_{m=0}^M\tilde{b}_{jn}\tilde{b}_{nm} 
          |{T_{(n-m)}^{\prime<}}{T_{(j-n)}^<} \epsilon_m \rangle + \sum_{n=0}^M\sum_{m=0}^M\tilde{b}_{jn}\tilde{c}_{nm} |{T_{(n-m)}^{\prime>}}{T_{(j-n)}^<} \epsilon_m \rangle  \nonumber \\ \fl+ \sum_{n=0}^M\sum_{m=0}^M\tilde{c}_{jn}\tilde{b}_{nm} 
          |{T_{(n-m)}^{\prime<}}{T_{(j-n)}^>} \epsilon_m \rangle + \sum_{n=0}^M\sum_{m=0}^M\tilde{c}_{jn}\tilde{c}_{nm} 
          |{T_{(n-m)}^>}^{(b)}{T_{(j-n)}^{\prime>}} \epsilon_m \rangle, 
        \end{eqnarray}
        where $T_{(n-m)}^{\prime}$=$T_0^{\prime}+\epsilon_n-\epsilon_m$
        and $T_{l}^{\prime}$ has the same meaning of  $T_{l}$
        but refers to the second scattered particle, i.e.
        electron $b$. As more electrons are scattered, the
        output state describing the system  involves more and more
        terms corresponding to the various scattering channels. For
        the sake of simplicity, here we only 
        describe the theoretical procedure used to calculate
        entanglement and decoherence in the case of the injection of a
        single carrier. In the
        subsection~\ref{currele},  the evaluation
        of  decoherence  and entanglement   due to the
        interaction with a large number of injected carriers (mimicking 
        an electric current) will be given for a specific case.

               In order to compute the   non-separability degree of the DQD system into the  
               product of single-particle states of the dots, that is the dot-dot entanglement, 
               we have to move from the description of the the DQD in terms of two-particle bound states to 
                the description in terms of single-particle bound states of the left and right dots.
          Thanks to the negligible Coulomb interaction
        between the two dots of  Figure~\ref{fig1}, the four lowest
         two-particle orbital 
        states $| \epsilon_n \rangle$ can be written in terms of 
        single-particle states $|E_l^{R}\rangle$ and
        $|E_m^{L}\rangle$, of the right and left dot, respectively  (Table~\ref{tab1}).
        \begin{table}
          \caption{\label{tab1} The table displays the scalar product $\langle E_n^L E_m^R | \epsilon_l \rangle$ for some values of $n,m,l$.}
          \begin{indented} 
            \lineup
          \item[] \begin{tabular}{ @{} c|c c c c } \br & $|E_0^L E_0^R
              \rangle$ & $|E_0^L E_1^R \rangle$ &$|E_1^L E_0^R
              \rangle$ &$|E_1^L E_1^R \rangle$ \\ \hline
              $|\epsilon_0 \rangle$ & 1 & 0 & 0 & 0 \\
              $|\epsilon_1\rangle$ & 0 & $-\frac{1}{\sqrt{2}}$ &  $-\frac{1}{\sqrt{2}}$ & 0 \\
              $|\epsilon_2\rangle$ & 0 & $-\frac{1}{\sqrt{2}}$ &  $\frac{1}{\sqrt{2}}$ & 0 \\
              $|\epsilon_3 \rangle$& 0 & 0 & 0& 1\\
            \end{tabular} 
          \end{indented}
        \end{table}
        In particular, $|\epsilon_0\rangle$
        is  the product of the ground states of the two dots,
        while  $|\epsilon_1\rangle$ and $|\epsilon_2\rangle$  are  degenerate,
       being each a
        linear superposition of $|E_0^L E_1^R \rangle$ and $|E_1^L
        E_0^R\rangle $, which represent one electron in the ground
        state of one dot and the other in the first excited state of
        the other. $|\epsilon_3\rangle$  corresponds to 
         the         first excited states of the         two  dots.
         States with two electrons in the same dot are included in our calculations but have always a negligible occupancy. This means that the coefficients
        $\tilde{b}_{jn}$,  and
        $\tilde{c}_{jn}$ with $j>3$ or $n>3$ on the right hand side of
        the equation~(\ref{phiout}) are 0. For our numerical calculations, we used the physical
        parameters of the GaAs material and the DQD potential reported in the caption of Figure~\ref{fig1}.

        In fact, we restrict our analysis to energies of the
        incoming particle which enable   up to four scattering
        channels, i.e.  the maximum value of $M$ in the
        expression~(\ref{phiout}) is  3. Under these
        assumptions a system of two qubits is obtained, in the sense
        that each  confined particle can be described in terms of two
        states: the ground $E_0^{L(R)}$ and first excited $E_1^{L(R)}$
        energy level of the left (right) QD, encoding the
        $|0_{L(R)}\rangle$ and $|1_{L(R)}\rangle$ states,
        respectively.  Thus, the three-particle quantum state of 
        equation~(\ref{phiout}) can be written as:
        \begin{eqnarray} \label{2threepart}\fl |\Phi_{OUT}\rangle
          = \tilde{b}_{j0} |T_j^< 0_{L}0_{R}\rangle -\frac{
            \tilde{b}_{j1}+
            \tilde{b}_{j2}}{\sqrt{2}}|T_{j-1}^<0_{L}1_{R} \rangle
          -\frac{ \tilde{b}_{j1}-
            \tilde{b}_{j2}}{\sqrt{2}}|T_{j-1}^<1_{L}0_{R} \rangle
          + \tilde{b}_{j3} |T_{j-3}^< 1_{L}1_{R} \rangle \nonumber \\
          \fl + \tilde{c}_{j0} |T_j^> 0_{L}0_{R}\rangle -\frac{
            \tilde{c}_{j1}+
            \tilde{c}_{j2}}{\sqrt{2}}|T_{j-1}^>0_{L}1_{R} \rangle
          -\frac{ \tilde{c}_{j1}-
            \tilde{c}_{j2}}{\sqrt{2}}|T_{j-1}^>1_{L}0_{R} \rangle +
          \tilde{c}_{j3} |T_{j-3}^> 1_{L}1_{R} \rangle
        \end{eqnarray}
        where $|T_{j-1}^>\rangle$=$|T_{j-2}^>\rangle$ and
        $|T_{j-1}^<\rangle$=$|T_{j-2}^<\rangle$, deriving from $E_0^L+
        E_1^R$=$E_1^L+ E_0^R$ and  energy conservation 
        has been taken into account.
	
        The decoherence undergone by the electrons confined in the
        dots can be interpreted in terms of the lack of knowledge of
        their quantum state  due to the
        interaction with the environment, namely the injected carrier~\cite{Peres}.  In other words, due to the
        coupling between the energy states stemming from the
        scattering event, the two-particle DQD cannot be
        described by a pure state anymore but becomes a statistical
        mixture.  A good measure of the degree of
        uncertainty about such a system and therefore of its loss of
        coherence, is given by the von Neumann entropy of the two
        particle reduced density matrix $\rho_r$, obtained by tracing
        the three-particle density matrix $\rho= |\Psi_{OUT}\rangle
        \langle \Psi_{OUT}|$ over the degrees of freedom $T_l^{>(<)}$
        of the scattered carrier~\cite{Peres}. After the first scattering, 
        the matrix
        representation of $\rho_r$ in the standard basis
        $\mathcal{B}=\{|0_{L}0_{R}\rangle,|0_{L}1_{R}\rangle,|1_{L}0_{R}\rangle,|1_{L}1_{R}\rangle
        \}$ reads
        \begin{equation}\label{RX}
          \rho_r= \left( \begin{array}{cccc}
              |\alpha|^2 & 0 & 0 & 0 \\
              0 &|\beta_+|^2 +|\gamma_+|^2&\beta_+\beta_-^{\ast}+\gamma_+\gamma_-^{\ast}& 0 \\
              0 & \beta_+^{\ast}\beta_-+\gamma_+^{\ast}\gamma_-&|\beta_-|^2 +|\gamma_-|^2& 0\\
              0 & 0& 0 &|\omega|^2
            \end{array}\right) ,
        \end{equation}
        where
        \begin{eqnarray}
          |\alpha|^2 &=&|\tilde{b}_{j0}|^2+|\tilde{c}_{j0}|^2 \nonumber \\
          \beta_{\pm}&=&\frac{\tilde{b}_{j1}\pm\tilde{b}_{j2}}{\sqrt{2}} \nonumber \\
          \gamma_{\pm}&=&\frac{\tilde{c}_{j1}\pm\tilde{c}_{j2}}{\sqrt{2}}
          \nonumber \\|\omega|^2 &=&|\tilde{b}_{j3}|^2+|\tilde{c}_{j3}|^2 .
        \end{eqnarray}
        In order to obtain the expression~(\ref{RX}), the
        orthogonality relations between the  states of the
        scattered carrier, $\langle T_i^{>(<)}|
        T_j^{>(<)}\rangle=\delta_{ij}$ and $\langle T_i^<|
        T_j^>\rangle=0 \quad \forall \,ij$, have been used.

        The decoherence $\xi$ can be evaluated by means of the von
        Neumann entropy as:
        \begin{equation}\label{deco} \xi=
          -\textrm{Tr}\left[\rho_r
            \ln{\rho_r}\right]= - |\alpha|^2 \ln{ | \alpha|^2}- \eta_+
          \ln{ \eta_+} -\eta_- \ln{ \eta_-} - |\omega|^2 \ln{
            |\omega|^2},
        \end{equation}
        where
        \begin{eqnarray}
          \eta_{\pm}=\frac{1}{2} \bigg(|\beta_+|^2 +|\gamma_+|^2+|\beta_-|^2 +|\gamma_-|^2 \nonumber \\
          \pm\sqrt{(|\beta_+|^2 +|\gamma_+|^2+|\beta_-|^2 +|\gamma_-|^2)^2-|\beta_+\gamma_--\beta_-\gamma_+|^2}\bigg).
        \end{eqnarray}
        It ranges from 0 to $\ln{3/2}$. For $\xi$=$0$ the two
        bound particles can be found in a single  energy level and this
        implies that no correlation is build up between them and the
        scattered carrier.  When the decoherence reaches its maximum, the DQD 
        is found in a statistical mixture of
        the  three allowed energies $\epsilon_0$, $\epsilon_1$, and
        $\epsilon_3$ with equal weight ($\epsilon_2$ is equal to $\epsilon_1$). This implies that the
        uncertainty about the system is maximum.

        The reduced density matrix $\rho_r$ describing the bound
        particles  is also used to evaluate the dot-dot entanglement
        through Wootters concurrence $C$.~\cite{Wootters} The latter
        is adopted to quantify the quantum correlations appearing
        between two qubits which cannot be described by a pure
        two-qubit state because of their coupling with an external
        environment, like in our scenario.          $C$ is obtained
        from the density matrix $\rho_r$ of the two-qubit system
        as~\cite{Wootters,Bellomo}:
        \begin{equation} \label{entag}
          C=\max{\left\{0,\sqrt{\lambda_1}-\sqrt{\lambda_2}-\sqrt{\lambda_3}-\sqrt{\lambda_4}\right\}},
        \end{equation}
        where $\lambda_i$ are the eigenvalues of the matrix
        $\zeta=\rho_r \left( \sigma_y^L\otimes
          \sigma_y^R\right)\rho_r^{\ast} \left( \sigma_y^L\otimes
          \sigma_y^R\right)$ arranged in decreasing order.  Here
        $\sigma_y^{L(R)}$ is the Pauli matrix $\bigg(
        \begin{array}{cc} 0 & -i \\ i & 0 \\ \end{array}\bigg)$ in the
        basis $\{|0_{L(R)}\rangle,|1_{L(R)}\rangle\}$, and
        $\rho_r^{\ast}$ describes the complex conjugation of $\rho_r$
        in the standard basis $\mathcal{B}$. The concurrence varies
        from $C$=0 for a disentangled state to $C$=1 for a maximally
        entangled state.

        The reduced density matrix $\rho_r$ given in  equation~(\ref{RX})
        shows an $\mathrm{X}$ structure, that is it contains non-zero
        elements only along the main diagonal and anti-diagonal. As
        shown in Ref.~\cite{Yu}, for such a class of density matrices
        the concurrence can be easily evaluated and in the case of
        $\rho_r$ it becomes:
        \begin{eqnarray} \label{enta1} C= 2\max{\left\{0,k\right\}},
        \end{eqnarray}
        where
        \begin{eqnarray} \label{enta2}
          k= \left|\beta_+\beta_-^{\ast}+\gamma_+\gamma_-^{\ast}\right| -|\alpha||\omega|=\nonumber \\
          \fl \frac{1}{2}
          \left|(\tilde{b}_{j1}+\tilde{b}_{j2})(\tilde{b}^{\ast}_{j1}-\tilde{b}^{\ast}_{j2})+
            (\tilde{c}_{j1}+\tilde{c}_{j2})(\tilde{c}^{\ast}_{j1}-\tilde{c}^{\ast}_{j2})\right|-\sqrt{\left(|\tilde{b}_{j0}|^2+|\tilde{c}_{j0}^2|
            \right)\left(|\tilde{b}_{j3}|^2+|\tilde{c}_{j3}^2|\right)}.
        \end{eqnarray}         
        $C$ is equal to 0 for $ |\alpha||\omega|\ge
        |\beta_+\beta_-^{\ast}+\gamma_+\gamma_-^{\ast}|$, while it
        reaches its maximum value 1 if and only if $|\beta_+\beta_-^{\ast}+\gamma_+\gamma_-^{\ast}|$=1 and $|\alpha|$=$|\omega|$=0 (and therefore the coefficients $\tilde{b}_{j0}$,
        $\tilde{c}_{j0}$, $\tilde{b}_{j3}$ and $\tilde{c}_{j3}$ vanish). In the latter case, the two-qubit system reduces to a
        Bell-like state $1/\sqrt{2}\bigg( |0_{L}1_{R} \rangle
        +e^{i\theta} |1_{L}0_{R} \rangle \bigg)$.

        \section{Results and discussion} \label{Results} Here we analyze our numerical results on  the decoherence $\xi$ undergone by the DQD  system and the
        dot-dot entanglement $C$. We stress again that the former corresponds to the entanglement of the DQD electrons with the transmitted/reflected one, while the latter is the concurrence between the two bound electrons.  In order to single-out the specific
        mechanisms leading  to the appearance of quantum correlations, the
        transmission and reflection spectra have been examined.
        \subsection{Scattering by a single carrier}

\begin{figure}[h]
          \begin{center}
            \includegraphics*[width=0.7\linewidth]{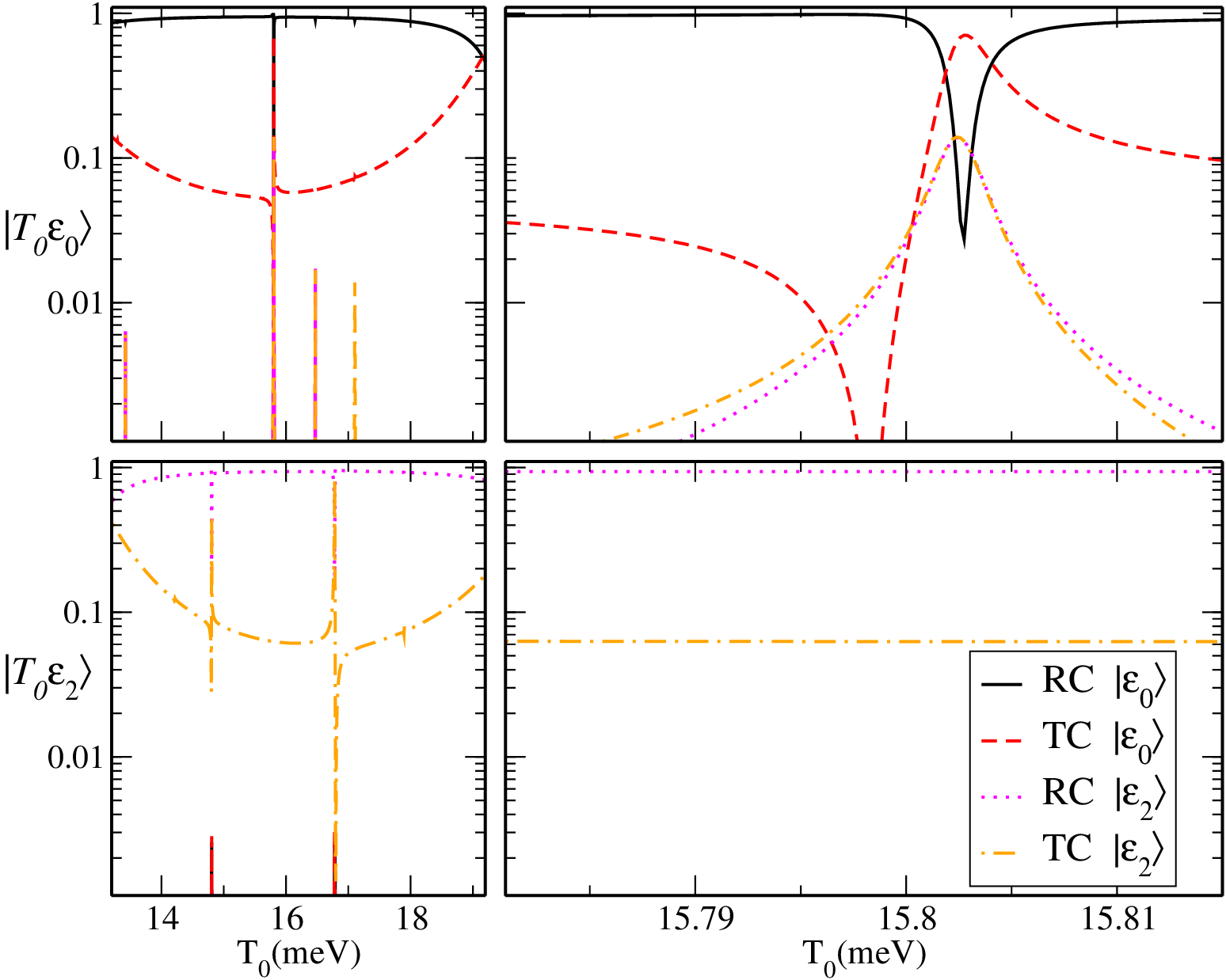}
            \caption{\label{fig2} Left panels: Modulus of the TC and
              RC of the channels corresponding to the DQD levels
              $|\epsilon_0 \rangle$ and $|\epsilon_2 \rangle$ 
              as a function of the initial kinetic energy of the
              incident electron $T_0$, ranging from 13.2 to 19.2 meV,
              for two different input states, namely $|T_0
              \epsilon_0\rangle$ (top), and $|T_0 \epsilon_2\rangle$
              (bottom): RC of the channel $|\epsilon_0\rangle$ (solid
              line), TC of the channel $|\epsilon_0\rangle$ (dashed
              line), RC of the channel $|\epsilon_2\rangle$ (dotted
              line), TC of the channel $|\epsilon_2\rangle$(dot-dashed
              line). 
             The modulus of TC and RC corresponding to the other two dot 
              states $|\epsilon_1\rangle$ and  $|\epsilon_3\rangle$, are zero in the region around the
              resonance energy $T_0$= 15.8 meV: for the sake of clarity, they have not been reported.
              Note that the sum of the moduli of the TC and RC is 1.
              Right panels: Modulus of the TC and RC of the channels
              $|\epsilon_0\rangle$ and $|\epsilon_2\rangle$ against
              $T_0$ close to the resonant energy $\bar{T}_0$ for the
              input states $|T_0 \epsilon_0\rangle$ (top), and $|T_0
              \epsilon_2\rangle$ (bottom).  }
             
          \end{center}
        \end{figure}

        The system has been solved for the potential
        profile $V_s(x)$ sketched in Figure~\ref{fig1} for various
        input states with  different 
        energy of the incoming carrier $T_0$.  In the top-left and
        bottom-left panels of Figure~\ref{fig2} we report the modulus
        of the transmission (TCs) and reflection coefficients (RCs) of
        the various scattering channels (that correspond  the energy levels of
        the two-particle DQDs system) for the input states $|T_0
        \epsilon_0\rangle$ and $|T_0 \epsilon_2\rangle$, respectively,
        when the incoming electron is injected with a kinetic  energy ranging
        from 13 to 19 meV.
        We recall  that the two-dot excited state
        $|\epsilon_2\rangle$ can be written in terms of the qubit
        states as $| \epsilon_2\rangle=1/\sqrt{2}\bigg( |0_{L}1_{R}
        \rangle - |1_{L}0_{R} \rangle \bigg)$, i.e. it is a Bell
        state.  The TCs and RCs are related to the coefficients
        $\tilde{b}_{jn}$ and $\tilde{c}_{jn}$ given in the
        expression~(\ref{phiout}).  For both the input configurations,
        sharp peaks are present only in small energy interval of the reflection
        and transmission spectra. It is worth noting that for
        different input states, such peaks appear at different values
        of $T_0$.

        To get a better insight into resonances, a zoom 
        of the modulus of TC and RC for  the various channels in the
        energy interval around the resonant condition $\bar{T}_0$=15.8
        meV is displayed in the right panels of the Figure~\ref{fig2}.
        While for the input state $|T_0 \epsilon_2\rangle$ no
        resonance is present in this range  and the sum of the moduli of TC and RC
        of the channel is equal to 1, for $|T_0
        \epsilon_0\rangle$ the  spectra
        show a number of sharp resonances (see the top-right panel of
        Figure~\ref{fig2}).  In particular the probabilities of
        finding the DQD in the state $|\epsilon_2\rangle$ (with the
        scattered carrier, either reflected or transmitted) shows a
        symmetric Lorentzian peak around $\bar{T}_0$, while the
        TCs and RCs of the scattering channel
         $|\epsilon_0\rangle$ exhibit a minimum.  Specifically,
        the RC presents a symmetric line-shape, and the TC an
        asymmetric one. This behavior clearly indicates that in the
        transmission and reflection spectra different kind of
        resonances appear, namely Breit-Wigner~\cite{Breit} and
        Fano~\cite{Fano}, respectively which have a strong connection into
        into the
        building up of quantum correlations, as noted
        elsewhere~\cite{busce,Hao,Hag}.
       The first ones, exhibiting symmetric Lorentzian peaks, stem
      from the coupling of a quasi bound state to the scattering states of the leads,
      while the second ones, characterized by asymmetric lineshapes, are present
      when two competing scattering mechanisms, a resonant one and  a nonresonant one, interfere, and are
     due to electron-electron correlation~\cite{bertoni,busce}.

\begin{figure}[h]
          \begin{center}
            \includegraphics*[width=0.7\linewidth]{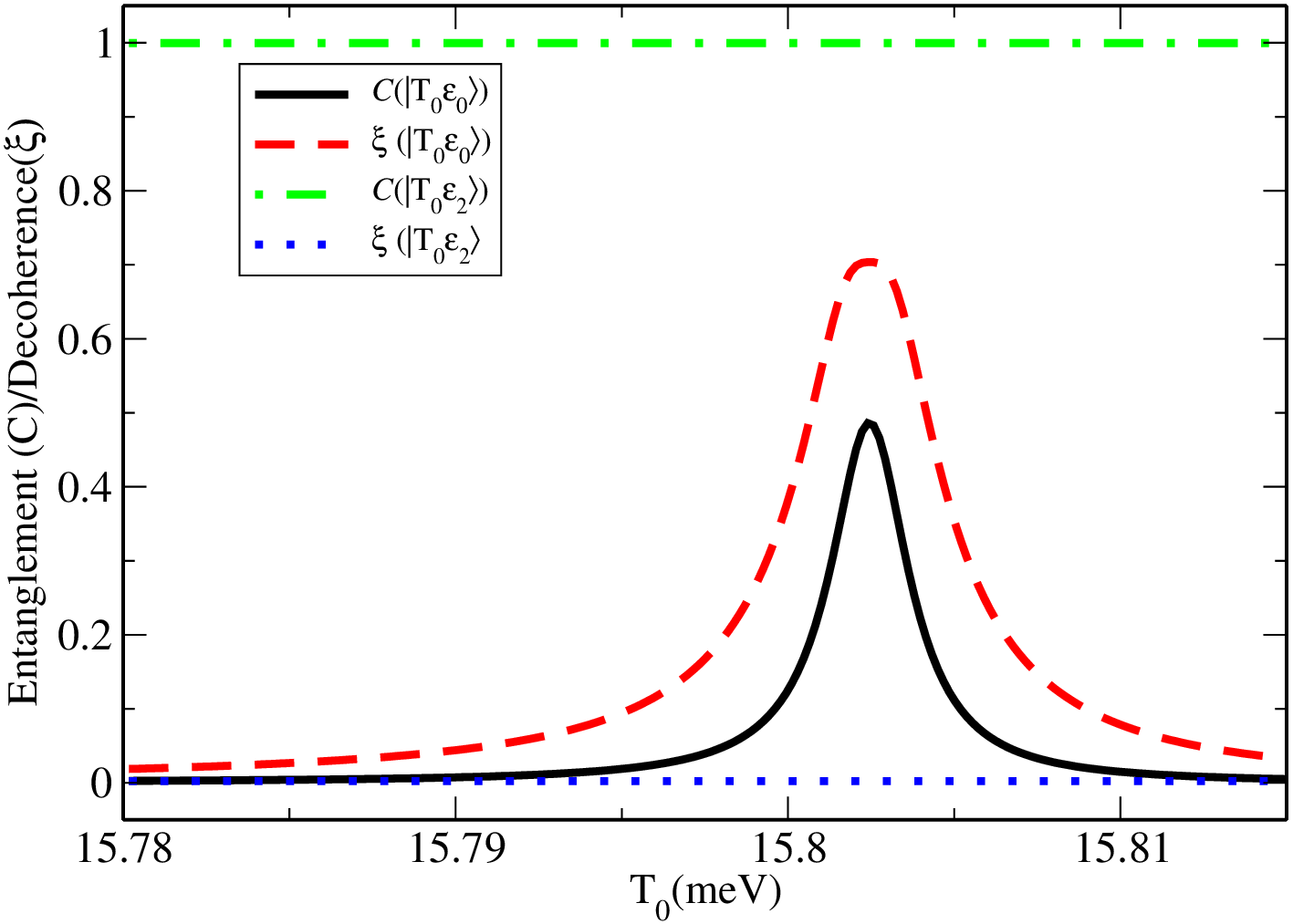}
            \caption{\label{fig3} Dot-dot entanglement $C$ and DQD decoherence
              $\xi$ against the initial kinetic energy $T_0$
              of the incident electron around the resonance condition
              $\bar{T}_0$, for two different input states: $C( |T_0
              \epsilon_0\rangle)$ (solid line), $\xi( |T_0
              \epsilon_0\rangle)$ (dashed line), $C( |T_0
              \epsilon_2\rangle)$ (dash-dotted line), and
              $\xi (|T_0 \epsilon_2\rangle)$ (dotted line).}
          \end{center}
        \end{figure}

        Scattering resonances are a signature of   peculiar decohering and
        entangling effects~\cite{busce}, as shown in Figure~\ref{fig3}, where the
        dependence of DQD decoherence $\xi$ and dot-dot entanglement $C$ are
        reported as a function of the initial energy of the incoming
        electron around $\bar{T}_0$ for the input states $|T_0
        \epsilon_0\rangle$ and $|T_0 \epsilon_2\rangle$.
        For $|T_0 \epsilon_2\rangle$, the decoherence is practically
        zero while the entanglement remains 1, as in the initial state.  In fact, as can be gathered by the
        behavior of TC and RC, the only coefficients not vanishing in
        equation~(\ref{phiout}) are $\tilde{b}_{22}$ and
        $\tilde{c}_{22}$, and the output state $|\Psi_{OUT}\rangle$ of
        expression~(\ref{2threepart}) reduces to
        \begin{equation}
          |\Psi_{OUT}\rangle=- \left(\tilde{b}_{22} |T^<_{0}\rangle+\tilde{c}_{22}|T^>_{0}\rangle \right) \frac{1}{\sqrt{2}} \left(|0_{L}1_{R} \rangle +|1_{L}0_{R} \rangle\right).
        \end{equation}
        This is the factorizable product of a single-particle state of
        the scattered carrier and  a two-qubit state describing the
        bound particles in the DQD, being the latter a Bell
        state.  This means that the entanglement between electrons
        confined in the dots maintains its maximum value and also that
        the quantum information about the DQD system is maximal since
        it is in a pure two-particle state.  Thus, the scattering event
        ``preserves'' the entanglement between the dots while the
        decoherence effects result to be negligible. Such a behavior,
        evidencing the existence of decoherence-free entangled states
        of two qubits, has been widely discussed in a number of works,
        which stressed the key role played by the symmetric coupling
        of the qubits with the environment into preserving their
        coherence~\cite{Zurek,Duan,Zana,Patra}

        When the input state is $|T_0 \epsilon_0\rangle$, both
        decoherence $\xi$ and entanglement $C$ show a maximum 
       where  RC and TC  of
        the channel $|\epsilon_2\rangle$ are resonant (see the top-right panel of         Figure~\ref{fig2}).  In particular $\xi$ reaches $\ln{2}$.
        Such a value is obtained when the DQD states are maximally
        coupled only to the energy levels $T_0$ and $T_{-2}$ of the
        scattered carrier and this occurs when the probabilities that
        the scattering leaves the bound particles in their ground
        $|\epsilon_0 \rangle$ or  excited $|\epsilon_2\rangle$
        state are equal. Thus the output state can be written as
        \begin{equation} \label{state2} \fl|\Psi_{OUT}\rangle=
          \left(\tilde{b}_{00}
            |T^<_{0}\rangle+\tilde{c}_{00}|T^>_{0}\rangle \right)
          |0_{L}0_{R} \rangle -\left(\tilde{b}_{02}
            |T^<_{-2}\rangle+\tilde{c}_{02}|T^>_{-2}\rangle \right)
          \frac{1}{\sqrt{2}} \left(|0_{L}1_{R} \rangle +|1_{L}0_{R}
            \rangle\right),
        \end{equation}       
        where $|\tilde{b}_{00}|^2+|\tilde{c}_{00}|^2$=$
        |\tilde{b}_{02}|^2+|\tilde{c}_{02}|^2\approx1/2$. In this
        case, from the expressions~(\ref{enta1}) and (\ref{enta2}), we
        observe that the value of the peak of the dot-dot entanglement
        is equal to 1/2, as shown in Figure~\ref{fig3}.  Here an important
        point to be stressed is  that quantum correlations are created  between
        the bound electrons,  even if their Coulomb interaction is
        negligible due to the large distance between the dots. In
        fact, even if these can be thought as totally decoupled
        subsystems, the external environment, i.e.  the scattered
        carrier, represents the interaction ``mediator'' 
       and it  represents the means to entangle
        them. In the literature, and on the basis of different
        physical mechanisms, the idea of an entanglement mediator has
        already been used in a number of theoretical and experimental
        models to produce bipartite entangled states
        ~\cite{Kazu,Browne,Compagno,Migliore,Costa}.

         \subsection{Scattering by an electric current}\label{currele}
         The above results indicate that, for two electrons each bound
         in the ground state of one of the dots, the interaction with a
         single incident carrier having a suitable 
         kinetic energy $\bar{T}_0$, excites the dots. Specifically,
         the scattering channel corresponding to $|\epsilon_2\rangle$,
         namely the Bell state describing the first DQD excited
         level, is activated and quantum correlations
         between the two dots appear, even if the bipartite
         entanglement production does not result to be maximal and
         immune to decohering effects. Indeed, the probability to
         excite the DQD is smaller than 1. This implies that the
         two dots cannot be described  in terms of the state
         $|\epsilon_2\rangle$ alone. Rather,  they are in a statistical mixture
         of ground and first-excited states.
         % evidencing also decoherence undergone by the system.
         
         On the other hand, the scattering between a carrier having
         kinetic energy $\bar{T}_0$ and the two electrons in the
         excited maximally-entangled state $|\epsilon_2\rangle$,
         leaves unchanged the DQD state, i.e. the entanglement is
         preserved and no decoherence effect appears. This behavior
         suggests that the maximum production of entanglement between
         dots set initially in their ground state, can be obtained as
         a consequence of the successive scatterings, one at a time,
         with  carriers injected with energy around $\bar{T}_0$. In
         fact, at each scattering event the probability of finding the
         DQD system in the excited state $|\epsilon_2\rangle $
         becomes larger, and  the amount of quantum
         correlations between the dots increases.  Such a sequence 
         of carrier injections corresponds to an electric current where all
         the electrons entering in the device have the same energy
         $\bar{T}_0$.From an experimental point of view, such a current
can be produced, for example, by using single-electron sources  such as,
electron pumps~\cite{vanWees},  resonant tunnelling diodes~\cite{Bjork},
or  systems consisting of a quantum dot connected to a conductor
via  a tunnel barrier~\cite{Feve}. All of these mechanisms,
enable to emit uncorrelated electrons in a given quantum state
with a specific  energy.

         In order to give a quantitative evaluation of the effect of
         an electric current on the DQD state, in
         \ref{Appendix} we have explicitely calculated the reduced
         density matrix $\rho^{(n)}_r$ describing the two dots after
         the injection of $n$ carriers. Its expression in the standard
         basis $\mathcal{B}$ is
         \begin{equation} \label{rhon} \rho_r^{(n)}= \left(
             \begin{array}{cccc}
               p_{00}^n & 0 & 0 & 0 \\
               0 & \frac{1- p_{00}^n}{2} & \frac{1- p_{00}^n}{2} & 0 \\
               0 &  \frac{1- p_{00}^n}{2} & \frac{1- p_{00}^n}{2} & 0\\
               0 & 0& 0 & 0
             \end{array}\right) ,
         \end{equation}
         where $p_{00}$=$|\tilde{b}_{00}|^2+|\tilde{c}_{00}|^2$,
         ranging from 0 to 1, is the probability that a scattering
         event leaves the QDs in the ground energy state when a
         carrier is injected with kinetic energy $T_0$. As stated
         before, for $T_0$= $\bar{T}_0$, $p_{00}$ is around 1/2.
         $\rho^{(n)}_r$ exhibits again an X structure and decoherence
         and entanglement of the system can be evaluated from the
         equations~(\ref{deco}) and (\ref{enta2}) by setting
         $|\alpha|^2$=$ p_{00}^n$, $\beta_+$=$\beta_-$=$\sqrt{(1-
           p_{00}^n)/2}$ and $\gamma_+$=$\gamma_-$=$\omega$=0. They 
         reads
         \begin{equation}
           \xi=-p_{00}^n\ln{p_{00}^n}-(1-p_{00}^n)\ln{(1-p_{00}^n)}\qquad  \textrm{and} \qquad C=1-p_{00}^n,
         \end{equation}
         respectively.  When $n$=0, i.e no scattering occurs, the
         expression~(\ref{rhon}) reduces to $\rho_r^{(0)}= |0_{L}0_{R} \rangle
         \langle 0_{L}0_{R}|$, which describes the input state where the DQD
         is in $|\epsilon_0\rangle = |0_{L}0_{R}\rangle $.  For $n$=1,
         $\rho_r^{(1)}$ is the reduced density matrix obtained from
         equation~(\ref{state2}) by tracing over $T^{<(>)}_i$. In the
         limit of large $n$, $\rho_r^{(n)}$ can be written as
         $\lim_{n\to \infty}\rho_r^{(n)}=\frac{1}{2} \left(|0_{L}1_{R}
           \rangle \langle 1_{R} 0_{L}| + |0_{L}1_{R} \rangle \langle
           0_{R} 1_{L}| + |1_{L}0_{R} \rangle \langle 1_{R} 0_{L}|
           +|1_{L}0_{R} \rangle \langle 0_{R} 1_{L}|\right)$ which
         corresponds to the Bell state of the two dots
         $\frac{1}{\sqrt{2}} \left(|0_{L}1_{R} \rangle +|1_{L}0_{R}
           \rangle\right)$ completely decoupled from the environment,
         with $\xi$=0 and $C$=1. That is, a current of independent electrons
         (with energy $\bar{T}_0$) entangles the two dots and does not create decoherence.

\begin{figure}[htbp]
	   \begin{center}
	     \includegraphics*[width=0.7\linewidth]{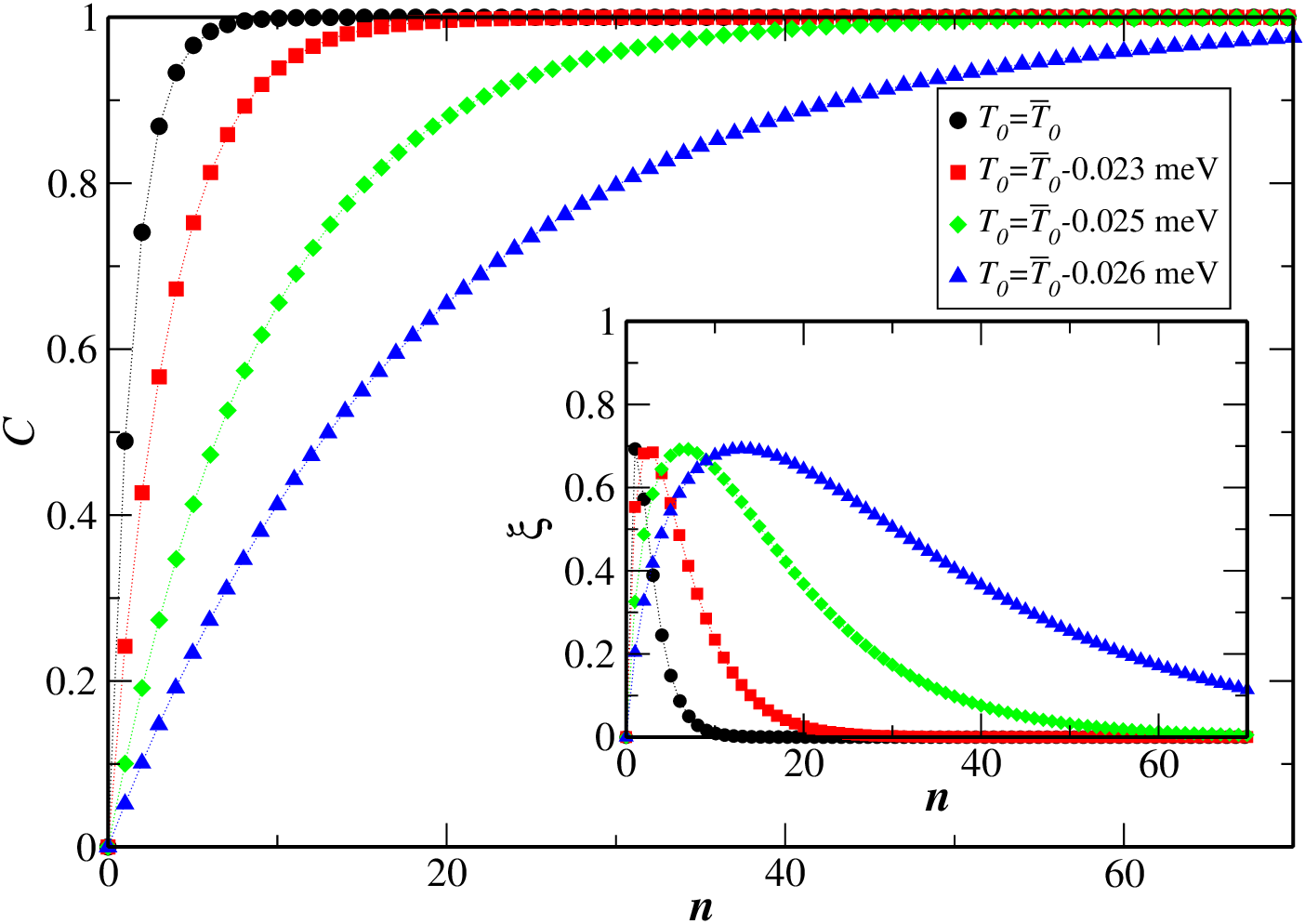}
	     \caption{\label{fig4} Dot-dot entanglement $C$ as a function of
               the number $n$ of the carriers injected into the device
               evaluated at four different values of the kinetic
               energy $T_0$ close to the resonance condition
               $\bar{T}_0$. As input condition, the bound particles
               occupy the ground state of the DQDs system.  All the
               curves tend to 1 for large values of $n$. The farther
               is the kinetic energy of each incident electron from
               $\bar{T}_0$, the slower the asymptotic value 1 is
               reached.  The inset displays the dependence of the
               DQD decoherence $\xi$ upon $n$, for the same four values of
               $T_0$.  }
	   \end{center}
	 \end{figure}

         Figure~\ref{fig4} displays the dependence of the entanglement
         upon the number $n$ of carriers entering in the device at
         different values of $T_0$ around the resonant energy
         $\bar{T}_0$.
         As shown in the inset of Figure~\ref{fig4}, we find  that
         a series of  scatterings 
         does not induce decoherence of the DQD  first-excited
         state  even for carriers injected with kinetic energies not exactly
         equal but close enough to $\bar{T}_0$.  This implies that maximally-entangled states of the
         DQD are produced as an effect of the flux of the charge
         carriers even if the energy of the incident electrons is not
         precisely the resonant one.  Specifically, the farther
         $T_0$ is from $\bar{T}_0$, the larger $n$ is needed to
         produce a Bell state decoupled from the environment. In fact,
         when the initial kinetic energy of the carriers gets away
         from the resonant one, the parameter $p_{00}$, acting as a
         convergence factor, increases and the injection of more
         carriers in the device is needed to build up the maximum
         amount of quantum correlations between the dots.  From the
         inset of Figure~\ref{fig4}, we also note that  the number $n$
         of electrons needed to have a vanishing decoherence (i.e. the DQD
         in a pure state) is lower for  $T_0$ closer to $\bar{T}_0$. 
         In particular, $\xi$ shows a
         maximum, whose value is about $\ln{2}$ when the interaction
         with the injected carriers reduces the state of the system to
         a statistical mixture  with equal weights of the ground
         $|\epsilon_0\rangle$ and the excited $|\epsilon_2\rangle$
         states.  This occurs for $p_{00}^n$=$2^{-1/n}$. As
         expected, the peak is at higher values of $n$ when the
         injection energy of the carriers is farther from $\bar{T}_0$
         and, as a consequence, $p_{00}$ is larger.

\begin{figure}[h]
	   \begin{center}
	     \includegraphics*[width=0.7\linewidth]{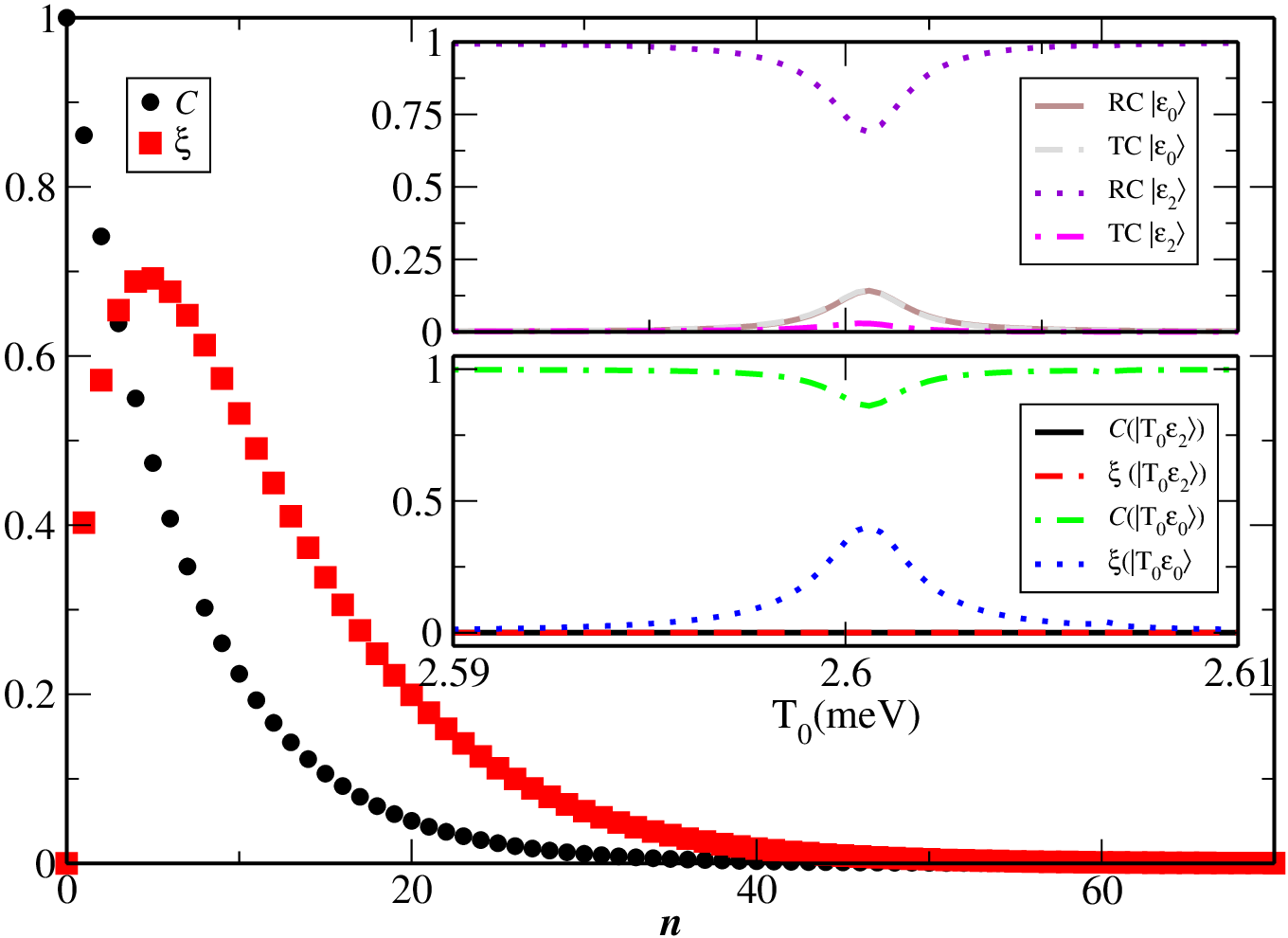}
	     \caption{\label{fig5} Dot-dot entanglement $C$ and DQD decoherence
               $\xi$ as a function of the number $n$ of the
               carriers injected into the device  at the
               resonant kinetic energy $T_0$=2.6 meV. As input
               condition, the bound particles occupy the first-excited
               state $|\epsilon_2 \rangle$ of the DQDs system
               corresponding to the Bell-state $\frac{1}{\sqrt {2}}
               \left(|0_{L}1 _{R} \rangle +|1_{L}0_{R}
                 \rangle\right)$.  Both $C$ and $\xi$ vanish at large
                $n$. The top inset displays the modulus of
               the TC and RC of the channels $|\epsilon_0\rangle$ and
               $|\epsilon_2\rangle$ of the DQD as a function of $T_0$
               around the resonant condition  when the input
               state of the total system is $|T_0 \epsilon_2\rangle$:
               RC of  channel $|\epsilon_0\rangle$ (solid line), TC
               of  channel $|\epsilon_0\rangle$ (dashed line), RC
               of  channel $|\epsilon_2\rangle$ (dotted line), TC
               of  channel $|\epsilon_2\rangle$ (dot-dashed line).
               The bottom inset shows the entanglement $C$ and the
               decoherence $\xi$ for the two input states $|T_0
               \epsilon_0\rangle$ and $|T_0 \epsilon_2\rangle)$: $C(
               |T_0 \epsilon_0\rangle)$ (solid line), $\xi( |T_0
               \epsilon_0\rangle)$ (dashed line), $C( |T_0
               \epsilon_2\rangle)$ (dash-dotted line), and $\xi( |T_0
               \epsilon_2\rangle)$ (dotted line). Note that the
               abscissa scale is the same in the top and bottom
               insets.}

	   \end{center}
	 \end{figure}

         In analogy  to the case of entanglement creation described above,
        a current
         of charge carriers injected at an appropriate energy
         can be a  means to disentangle the DQD
         prepared in the Bell state $|\epsilon_2 \rangle\frac{1}{\sqrt
           {2}} \left(|0_{L}1 _{R} \rangle +|1_{L}0_{R}
           \rangle\right)$.  In order to show this, Figure~\ref{fig5} displays the
         disentanglement effect for electrons injected with kinetic
         energy $T_0$ around 2.6 meV and scattered by the bound
         particles of the DQDs.
         For such a low kinetic energy, the scattering by a single
         carrier leaves unaltered the DQDs system when the bound
         electrons are in the ground state $|\epsilon_0\rangle$. In
         fact $T_0$ is smaller than the energy necessary to excite
        the  dots.  This means that the the scattered
         carrier has not been coupled, via Coulomb interaction, to the
         bound particles, which remain maximally disentangled (see the
         bottom inset of the Figure~\ref{fig5}). On the other hand,
         when the input state of the total system is $|T_0
         \epsilon_2\rangle$ the dots can relax. In fact  the scattering
         channels corresponding to $|\epsilon _0\rangle$ show a peak
         in the transmission and reflection spectra (as shown in the
         top inset of Figure~\ref{fig5}) thus leading to the
         appearance of  decoherence and entanglement.

         By applying the approach adopted above for building up 
         maximum entanglement  between the bound
         electrons, we find that the scattering by a current of charge carriers
         with  energy around $T_0=2.6$ meV is able to disentangle
         completely the quantum dots without introducing any decoherence. 
        Specifically, after  a larger number $n$ of  scattered carriers,
         the bound electrons practically occupy the ground
         state of the DQD system: this means $C=0$ and $\xi=0$, as
         reported in Figure~\ref{fig5}.

         \section{Conclusions}\label{Conclu}
         The coherent manipulation of electron states is the key
         ingredient for implementing  qubits using  charge or orbital
        states of DQD
         nanostructures. Indeed, it implies the controlled production
         or destruction, the manipulation and detection of
         entanglement between the above  states.  In this spirit, various
         proposals to produce bipartite entangled state have been
         advanced on the basis of the physical mechanisms requiring
         two-particle scattering, such as the direct interaction
         between two electrons~\cite{Oliver,Lopez, busce,busce3,
           Schomerus}.  In this work,  we have investigated the
         appearance of quantum correlations between the
        two electrons of a GaAs DQD,
          as a consequence of the Coulomb scattering by one or more 
         charge carriers injected from a lead. We examined the
         scattering event in a three-particle regime (the two
         electrons trapped in the DQDs and the passing carrier,
        explicitely considered indistinguishable), that is, a
         carrier is supposed to enter in the scattering region only
         after the previous one has already left.
         Furthermore, the two dots are taken distant enough so
          that the Coulomb repulsion between
         the two bound electrons is practically negligible. 
         Therefore, unlike other approaches~\cite{Lopez, busce}, the scattered carrier
         represents the entanglement ``mediator'', that is, it provides
         the indirect interaction between the particles needed to
         entangle them. From this point of view, various scheme where entanglement between 
         distant particles is produced though their scattering
         by mobile mediators are present in literature~\cite{Costa,Cic2}. Unlike our model,
         there the quantum correlations are built among the spin degrees
         of the freedom of the particles.

         A proper tuning of the carrier energy and the DQD geometry
         reduces the system here examined to a simple two-qubit model
         coupled to the external degrees of freedom by  the incident
         electrons. Here, the dots have not been considered as
         point-like systems~\cite{Wu,Voro,Cao,Openov,Li} but their
         effective spatial dimensions are explicitely taken into
         account in the calculation. Indeed, the knowledge of the
         electron spatial  wavefunctions corresponding to eigenstates of
         the DQD  is needed to obtain the few-particle scattering
         states. To this aim, a time-independent approach based on the
        QTBM  has been
         used~\cite{lent,bertoni,busce}. Its solution gives the
         reflection and transmission  amplitude of each
         scattering channel as a function of the initial energy of
         the incoming electron. Such an approach permits to analyze the
         relation between the  resonances in the transmission and reflection
         spectra and the  appearance of quantum correlations between the
         particles, as already pointed  out elsewhere~\cite{Lopez, busce,Kazu}. All
         the travelling components of the scattered carrier, both
         reflected and transmitted, have been used to evaluate the
         creation of the entanglement  between the dots, together with  their
         decoherence.

         Our numerical simulations show that as a consequence of the
         scattering  between an electron  injected with a
         suitable energy and two electrons bound in the ground state
         of the DQD system, the latter can be excited, ending up in 
         an entangled state of the constituent dots. This process
         leads to the appearance of resonance peaks
         and dips in transmission and reflection spectra
         of  the first-excited and ground scattering
         channels, respectively.  A side-effect of such a scattering is the loss
        of  quantum coherence of the DQD as whole due to its coupling to the
          scattered carrier. The condition of
         maximum entanglement between the two dots
         is reached  when the bound electrons
         are fully raised to the two-particle first-excited level of
         the DQD system (which corresponds to a Bell-state formed
         with the single-particle ground and first-excited states of
         the two dots).  In this case, the DQD decoherence
         is zero, since a single output channel is possible. However, a single           collision is not able
         to fully excite  the dots. We found that, in order  to build up the
         maximum amount of quantum correlation between them, a repeated 
          injection of  charge
         carriers, that is an electric current, is needed. Indeed, at
         each scattering event the excitation probability of the dots
         increases until it reaches asymptotically 1, which means that
         a Bell state is obtained, fully decoupled from the degrees of
         freedom of the scattered carriers.  In other words, the
         Coulomb interaction between an electric current and two
         electrons bound in the ground state of a DQD structure
         allows for the maximum entanglement production while the
         decoherence effects on the system vanish. This is in
         agreement with the procedures adopted in other
         works~\cite{Kazu,Browne,Compagno,Migliore,Costa}, where the
         entangling schemes are based on the successive interactions
         of a mediator with the qubits. However, in our
         scheme, the indirect coupling of the two dots  due to
         interaction with the scattered carriers, can produce
         disentangling effects, as well.  Indeed, a proper tuning of
         the electric current makes the DQD, initially in a Bell
         state to relax into the ground state, with no quantum
         correlations. Also in this case the process results
         to be robust against decoherence.

         Finally, the results here reported show how a  suitable
         electron current, where all the carriers have almost a given
         kinetic energy, permits to switch coherently on and off
         the entanglement between the dots of a DQD structure. Although
         several interaction mechanisms, as electron-phonon coupling
         can lead to the loss of quantum coherence of the DQD
         in a real experimental setup, we showed
         that interaction with the mediator electrons
        does not generate entanglement with the leads. Thus, no intrinsic decoherence is implied.

         \appendix
         \section{Evaluation of the reduced density matrix
           $\rho_r^{(n)}$} \label{Appendix}

         Here we shall give the explicit derivation  of the reduced
         density matrix of the two dots (see equation (\ref{rhon})), 
         initially taken in their
         ground state, once $n$ carriers  injected with kinetic
         energy $T_{0}$ close to $\bar{T_{0}}$ have been scattered.
          In order to simplify
         the calculation, the basis
         $\mathcal{C}=\{|\epsilon_0\rangle,|\epsilon_1\rangle,
         |\epsilon_2\rangle,|\epsilon_3\rangle\}$ of the DQD eigenstates
         will be used. Once obtained the
         reduced density operator ${\rho_r^{(n)}}^{\prime}$ in the
         $\mathcal{C}$ basis, its expression ${\rho_r^{(n)}}$ in terms
         of the $\mathcal{B}$ is 
         straightforward (see Table~\ref{tab1}).

         The output three-particle state of equation~(\ref{state2}),
         stemming from the scattering between one carrier injected in
         the device with $T_{0}$=$\bar{T_{0}}$ and the two electrons
         in the ground state of the DQD, can be written as
         \begin{equation}
           \fl|{\Psi_{OUT}^{(1)}}^{\prime}\rangle=
           \left(\tilde{b}_{00} |{T^<_{0}}^{(1)}\rangle+\tilde{c}_{00}|{T^>_{0}}^{(1)}\rangle \right)
           |\epsilon_0 \rangle +\left(\tilde{b}_{02}
             |{T^<_{-2}}^{(1)}\rangle+\tilde{c}_{02}|{T^>_{-2}}^{(1)}\rangle \right)|\epsilon_2\rangle ,
         \end{equation}
         where the superscript (1) means 1 carrier injected, and the reduced density
         matrix of the DQD system can be obtained by tracing
         $|{\Psi_{OUT}^{(1)}}^{\prime}\rangle \langle
         {\Psi_{OUT}^{(1)}}^{\prime}|$ over the degrees of freedom
         ${T^{>(<)}_{i}}^{(1)}$ of the scattered carrier
         \begin{equation}
           { \rho_r^{(1)}}^{\prime}= \left( \begin{array}{cccc}
               p_{00} & 0 & 0 & 0 \\
               0 & 0 & 0 & 0 \\
               0 &  0 & 1- p_{00}& 0\\
               0 & 0& 0 & 0
             \end{array}\right).
         \end{equation}
         When a second carrier is injected, after the exit of the
         previous one from the scattering region, the new output state
         is
         \begin{eqnarray} \label{psi2} \fl
           |{\Psi^{(2)}_{OUT}}^{\prime}\rangle = \left(\tilde{b}_{00}
             |{T^<_{0}}^{(2)}\rangle+\tilde{c}_{00}|{T^>_{0}}^{(2)}\rangle
           \right) \left(\tilde{b}_{00}
             |{T^<_{0}}^{(1)}\rangle+\tilde{c}_{00}|{T^>_{0}}^{(1)}\rangle
           \right)
           |\epsilon_0 \rangle +\nonumber \\
           \left(\tilde{b}_{02}
             |{T^<_{-2}}^{(2)}\rangle+\tilde{c}_{02}|{T^>_{-2}}^{(2)}\rangle
           \right) \left(\tilde{b}_{00}
             |{T^<_{0}}^{(1)}\rangle+\tilde{c}_{00}|{T^>_{0}}^{(1)}\rangle
           \right) |\epsilon_2 \rangle +
           \nonumber \\
           \left(\tilde{b}_{22}
             |{T^<_{0}}^{(2)}\rangle+\tilde{c}_{22}|{T^>_{0}}^{(2)}\rangle
           \right) \left(\tilde{b}_{02}
             |{T^<_{-2}}^{(1)}\rangle+\tilde{c}_{02}|{T^>_{-2}}^{(1)}\rangle
           \right)|\epsilon_2\rangle .
         \end{eqnarray} As stressed  in Section~\ref{Results}, when the DQD             is in $| \epsilon_2 \rangle$, the scattering event does not produce  the relaxation of           the dots
         which remain in the excited energy level. This means that in the above expression
         $|\tilde{b}_{22}|^2+|\tilde{c}_{22}|^2=1$.  The  reduced density matrix of the DQD 
         computed from the three-particle state of equation~(\ref{psi2}) is
         \begin{equation}
           {\rho_r^{(2)}}^{\prime}= \left( \begin{array}{cccc}
               p_{00}^2 & 0 & 0 & 0 \\
               0 & 0 & 0 & 0 \\
               0 &  0 & (1- p_{00}^2) & 0\\
               0 & 0& 0 & 0
             \end{array}\right) ,
         \end{equation}
         where 
         $p_{00}$=$|\tilde{b}_{00}|^2+|\tilde{c}_{00}|^2$=$1-|\tilde{b}_{02}|^2-|\tilde{c}_{02}|^2$
         have been used. For the case of $n$ scattered particles we get
         \begin{equation} \label{casen} {\rho_r^{(n)}}^{\prime}=
           \left(
             \begin{array}{cccc}
               p_{00}^n & 0 & 0 & 0 \\
               0 & 0 & 0 & 0 \\
               0 &  0 & (1- p_{00}^n) & 0\\
               0 & 0& 0 & 0
             \end{array}\right) ,
         \end{equation}
         as derived by induction in the following.  Assume that
         expression (\ref{casen}) is true for $n$. This implies that $
         {\rho_r^{(n)}}^{\prime}= p_{00}^n |\epsilon_0\rangle \langle
         \epsilon_0| +(1- p_{00}^n)|\epsilon_2\rangle \langle
         \epsilon_2|$. After the injection of the $(n+1)$-th carrier,
         the density matrix
         $\rho^{\prime}(\epsilon_i,\epsilon_j,{T^{<(>)}_l}^{(n+1)},{T^{<(>)}_m}^{(n+1)})
         $ describing the total system can be evaluated from
         ${\rho_r^{(n)}}^{\prime}$: 
         \begin{eqnarray}
           \fl  \rho^{\prime}(\epsilon_i,\epsilon_j,{T^{<(>)}_j}^{(n+1)},{T^{<(>)}_m}^{(n+1)})=\nonumber \\  p_{00}^n \bigg[ \left(\tilde{b}_{00} |{T^<_{0}}^{(n+1)}\rangle+\tilde{c}_{00}|{T^>_{0}}^{(n+1)}\rangle \right)            |\epsilon_0 \rangle \langle \epsilon_0| \left(\tilde{b}_{00}^{\ast}              \langle {T^<_{0}}^{(n+1)}|+\tilde{c}_{00}^{\ast} \langle {T^>_{0}}^{(n+1)}| \right)
           + \nonumber \\
           \left(\tilde{b}_{00} |{T^<_{0}}^{(n+1)}\rangle+\tilde{c}_{00}|{T^>_{0}}^{(n+1)}\rangle \right)            |\epsilon_0 \rangle \langle \epsilon_2| \left(\tilde{b}_{02}^{\ast}              \langle {T^<_{-2}}^{(n+1)}|+\tilde{c}_{02}^{\ast} \langle {T^>_{-2}}^{(n+1)}| \right)  +\nonumber \\
           \left(\tilde{b}_{02} |{T^<_{-2}}^{(n+1)}\rangle+\tilde{c}_{02}|{T^>_{-2}}^{(n+1)}\rangle \right)            |\epsilon_2 \rangle \langle \epsilon_0| \left(\tilde{b}_{00}^{\ast}              \langle {T^<_{0}}^{(n+1)}|+\tilde{c}_{00}^{\ast} \langle {T^>_{0}}^{(n+1)}| \right) + \nonumber \\
           \left(\tilde{b}_{02} |{T^<_{-2}}^{(n+1)}\rangle+\tilde{c}_{02}|{T^>_{-2}}^{(n+1)}\rangle \right)            |\epsilon_2 \rangle \langle \epsilon_2| \left(\tilde{b}_{02}^{\ast}              \langle {T^<_{-2}}^{(n+1)}|+\tilde{c}_{02}^{\ast} \langle {T^>_{-2}}^{(n+1)}| \right) \bigg]+ \nonumber \\  \fl (1-p_{00}^n )\left(\tilde{b}_{22}
             |{T^<_{0}}^{(n+1)}\rangle+\tilde{c}_{22}|{T^>_{0}}^{(n+1)}\rangle \right)|\epsilon_2\rangle \langle \epsilon_2| \left(\tilde{b}_{22}^{\ast}              \langle {T^<_{0}}^{(n+1)}|+\tilde{c}_{22}^{\ast} \langle {T^>_{0}}^{(n+1)}| \right).
         \end{eqnarray}
         By tracing 
         $\rho^{\prime}(\epsilon_i,\epsilon_j,{T^{<(>)}_j}^{(n+1)},{T^{<(>)}_m}^{(n+1)})
         $ over the degrees of freedom ${T^{<(>)}_j}^{(n+1)}$ of the
         carrier, one obtains the reduced density matrix
         ${\rho_r^{(n+1)}}^{\prime}$ of the DQD scattered by $n+1$
         electrons. Its expression reads
         \begin{eqnarray}
           \fl  {\rho_r^{(n+1)}}^{\prime}=  p_{00}^n \bigg(|b_{00}|^2+|c_{00}|^2\bigg)|\epsilon_0\rangle \langle \epsilon_0|
           +\bigg(p_{00}^n (|b_{02}|^2+|c_{02}|^2)+(1- p_{00}^n)(|b_{22}|^2+|c_{22}|^2)\bigg)|\epsilon_2\rangle \langle \epsilon_2| = \nonumber \\
           \fl p_{00}^{n+1} |\epsilon_0\rangle \langle \epsilon_0| +(1- p_{00}^{n+1})|\epsilon_2\rangle \langle \epsilon_2|.
         \end{eqnarray}
         Thus the expression~(\ref{casen}) is true for $n+1$. 

        Finally,
         the unitary transformation of Table~\ref{tab1} can be applied to
         ${\rho_r^{(n)}}^{\prime}$ in order to obtain the reduced density
         matrix of the two electrons bound in the DQD given in
         equation~(\ref{rhon}).

	\end{document}